\documentstyle[11pt,epsfig]{article}
\def\be{\begin{eqnarray}}
\def\ee{\end{eqnarray}}
\def\ben{\begin{equation}}
\def\een{\end{equation}}
\def\bi{\bibitem}

\def\calL{{\cal L}}

\def\prl{Phys. Rev. Lett.}\def\pr{Phys. Rev.}\def\np{Nucl. Phys.}
\def\pl{Phys. Lett.}
\def\J#1#2#3#4{ {#1} {\bf #2} {#3} {(#4)} }

\def\PLB{Phys. Lett. B}
\def\NPB{Nucl. Phys. B}
\def\PRC{Phys. Rev. C}

\def\del{\partial}

\def\roughly#1{\mathrel{\raise.3ex\hbox{$#1$\kern-.75em%
\lower1ex\hbox{$\sim$}}}}
\def\gsim{\roughly>}

\def\ve{\varepsilon}

\def\He#1{{}^{#1}\mbox{He}}

\def\itt{\indent\indent}
\def\E{{\cal E}}

\def\Jb{\mbox{\boldmath$ J$}}

\def\jb{\mbox{\boldmath$ j$}}
\def\pb{\mbox{\boldmath$ p$}}\def\qb{\mbox{\boldmath $q$}}

 \def\vb{\mbox{\boldmath$ v$}}

\def\Bq{\mbox{\boldmath $q$}}
\def\Bk{\mbox{\boldmath$ k$}}

\def\qbhat{\mbox{\boldmath $\hat{q}$}}\def\taub{\mbox{\boldmath $\tau$}}

\def\kbhat{\mbox{\boldmath $\hat{k}$}}

\def\gammab{\mbox{\boldmath $\gamma$}}

\def\sigmab{\mbox{\boldmath $\sigma$}}

\def\Bsigma{\mbox{\boldmath $\sigma$}}

\def\A0{A_0}
\def\bq{\begin{equation}}
\def\eq{\end{equation}}

\renewcommand{\thefootnote}{\fnsymbol{footnote}}
\begin{document}
\begin{titlepage}
\begin{center}
\hfill {KIAS-P00040}

 \vskip 2cm {\Large \bf Effective Field Theories, Landau-Migdal
 Fermi-Liquid Theory\\ and Effective Chiral Lagrangians for Nuclear
 Matter\footnote{
 Dedicated to the memory of A.B.~Migdal on
 the 90th anniversary of his birthday}}
  \vskip 1cm
   {{\large Mannque Rho$^{a,b}$} } 

{\it (a) School of Physics, Korea Institute for Advanced Study,
Seoul 133-791, Korea}

{\it (b) Service de Physique Th\'eorique, CE Saclay, 91191
Gif-sur-Yvette, France}

\end{center}

\vskip 1cm

\centerline{(\today)}
 \vskip 1cm

\centerline{\bf Abstract} \vskip 0.5cm

We reinterpret Landau-Migdal Fermi-liquid theory of nuclear matter
as an effective chiral field theory with a Fermi surface. The
effective field theory is formulated in terms of a chiral
Lagrangian with its {\it mass and coupling parameters} scaling a
la Brown-Rho and with the Landau-Migdal parameters identified as
the fixed points of the field theory. We show how this mapping
works out for response functions to the EM vector current and then
using the same reasoning, make prediction on nuclear axial
current, in particular on the enhanced axial-charge transitions in
heavy nuclei. We speculate on how to extrapolate the resulting
theory which appears to be sound both theoretically and
empirically up to normal nuclear matter density $\rho_0$, to
hitherto unexplored higher density regime relevant to
relativistic heavy-ion processes and to cold compact (neutron)
stars.
\end{titlepage}
\newpage

\renewcommand{\thefootnote}{\arabic{footnote}}
\setcounter{footnote}{0}

\section{Introduction}
\itt
 In a recent beautiful development~\cite{shankar}, Landau's Fermi
liquid theory has been re-formulated as a modern effective field
theory with the Fermi liquid state identified as a stable fixed
point.
This theory represents an effective field theory which is as
beautiful as chiral Lagrangian field theory for low-energy pionic
interactions. It is then most natural that Migdal's theory of
nuclear matter~\cite{migdal} which is based on Landau Fermi liquid
theory can also be formulated as an effective field theory. We
dedicate this note, which is based on recent
work~\cite{FRSall,songPR}, as a tribute to Migdal on the occasion
of his 90th anniversary.
\section{Effective Field Theory}
\itt
 Effective field theories enter the nuclear physics domain in two
different ways. One is to make precise predictions for certain
processes involving few-nucleon systems that are connected with
fundamental issues of physics. This is often called for to answer
questions of fundamental nature in other areas of physics such as
astrophysics or particle physics~\cite{taipei}. The other
-- which is our objective here -- is to be able to extrapolate the
knowledge available in normal conditions beyond the normal nuclear
matter regime into a high temperature or high density regime that
will be the focus of experimental efforts in the coming years. In
making the extrapolation, the usual quantum mechanical many-body
approach lacks the necessary versatility and field theoretic
approaches anchored in quantum chromodynamics will be required.
Migdal's
formulation of Fermi liquid for nuclear matter has proven powerful
at least up to normal nuclear matter density, and has even led to a
variety of predictions of phenomena that might take place in
extreme conditions~\cite{migdalpi}. In its original form, however,
it is somewhat limited in its scope if one wishes to extrapolate to
extreme conditions, where QCD phase changes may be induced. Such
densities are expected in upcoming laboratories and probably exist
in neutron stars interiors. In this note, we wish to discuss our
recent attempt to formulate Landau-Migdal theory of nuclear matter
in a modern effective field theory framework. Such a framework,
which offers the possibility of extrapolation to extreme
conditions, has been quite successful in such different fields as
condensed matter and high-energy physics.
\subsection{Effective field theory (EFT) for light nuclei}
\itt
Before going into our main topic of dense matter, we briefly
summarize the status of effective field theories in few-nucleon
systems. Here the setting for an EFT is straightforward.

The objectives there are essentially two-fold. One is to derive the
nucleon-nucleon interactions -- which are fairly well understood
from phenomenological approaches -- from fundamental principles.
The basic question is: Can all two-nucleon systems, viz,
nucleon-nucleon scattering at low energy and bound states (e.g.,
the deuteron) be understood in the framework of an effective field
theory? This is an old question, which stimulated by the work of
Weinberg~\cite{weinberg}, recently became the focus of intense
activities in many theoretical communities. The status of the field
is comprehensively summarized in the proceedings of two recent
INT-Caltech workshops~\cite{INTproceed}. The original Weinberg
approach had certain {\it apparent} inconsistency in the power
counting invoked for a systematic calculation but this problem can
be readily resolved as shown by the INT-Caltech
collaboration~\cite{KSW}. In this work the notion of ``power
divergence subtraction" was introduced into the dimensional
regularization. This enables one to handle the anomalous length
scale that appears when a quasi-bound state is near by in a more
straightforward manner. We now know that when done correctly, the
two schemes (i.e., Weinberg's and the INT/Seattle-Caltech scheme),
are essentially equivalent in the treatment of low-energy
two-nucleon interactions.
Although they may differ in specific details, the two schemes
reproduce the low-energy observables thus far studied with equal
quality.

The other objective is to exploit the power of effective field
theories in making {\it bona-fide predictions} for processes which
cannot be accessed by standard nuclear physics methods. Examples
that have been discussed recently are the asymmetry observables in
the polarized $np$ capture~\cite{PKMR-np}
\be
\vec{n}+\vec{p}\rightarrow d+\gamma\label{polnp}
 \ee
 and the solar {\it hep} process~\cite{PKMR-hep}
 \be
p+\He3 \rightarrow \He4 + e^+ + \nu_e. \label{hep}
 \ee The first process (\ref{polnp}) has been studied theoretically
in a variety of different methods~\cite{PKMR-np,CRS-np} and is
being measured~\cite{mueller}. The second process (\ref{hep}) has
been recently measured in the Super-Kamiokande
experiment~\cite{superK} and has generated a lot of excitement
among theorists~\cite{bahcall}.  It turns out rather remarkably
that these two processes complement each other in providing a
theoretical strategy to overcome a non-trivial obstacle on the way
to a parameter-free calculation.

Now, in order to increase the predictive power in general and to
facilitate accurate calculations of the above processes, a hybrid
version of EFT (called MEEFT or ``more effective EFT") was launched
by Park, Kubodera, Min and Rho~\cite{PKMR,PKMR-astro,PKMR-hep}.
This approach, which combines the sophisticated standard nuclear
physics approach with chiral perturbation theory, turns out to be
far more powerful and robust than naively expected. Within this
scheme one can {\it actually} make reliable calculations of
observables that cannot be obtained by other methods. Of equal
importance is the fact that such predictions can be confronted with
data. Thus, the validity of this approach will be settled by
experiment in the near future. The accuracy with which such
predictions can be made is assessed in \cite{PKMR-hep}.

\subsection{EFT for heavy nuclei and nuclear matter}
\itt
 In both cases mentioned above addressing low-density systems, the
effective Lagrangians are defined at zero density and the relevant
fluctuations are made on top of the zero-density vacuum which is
accessible by various QCD analyses, treating the matter density as
an external perturbation. In a dense medium, the situation is
expected to be quite different. While in the light systems the
parameters that figure in the effective Lagrangian are in principle
derivable from QCD (perhaps on a lattice) or more often from
experimental data, this is not the case in a dense medium. Deriving
an EFT for dense matter from QCD is probably of similar difficulty
as deriving the Hubbard model from QED. The best one can do is to
start with a Lagrangian defined at zero density and go up in
density. Unfortunately this will be limited to low density and
cannot be pushed to high enough density to be useful in the regime
we would like to explore.

In this note we circumvent the difficulty of deriving such a theory
directly. Rather, we construct an effective chiral Lagrangian field
theory that {\it maps} onto an established many-body theory,
specifically Landau-Migdal's Fermi-liquid theory and then
extrapolate that field theory to the regime of higher density. This
is certainly in accordance with the original spirit of
Landau-Migdal theory though it is not clear that such a scheme will
work in all density regimes. We can only say that up to now there
is no evidence against the scheme. For a recent review, see
\cite{songPR}.
\section{Nuclear Matter as a Fermi Liquid Fixed Point}
\subsection{Chiral liquid}
 \itt How to obtain a realistic description of nuclear matter from
an effective Lagrangian anchored in the fundamentals of QCD is very
much an open problem at the moment. There are however several
models available. One of them, the skyrmion with an infinite baryon
number is yet to be confronted with Nature. The skyrmion is a
soliton solution of a Skyrme-type Lagrangian, which is an
approximate Lagrangian for QCD at infinite number of colors
$N_c=\infty$. Because the mathematical structure of this model is
not very well known at the moment, only very little information can
be extracted from it.

Another model is the non-topological soliton picture proposed in
an embryonic form sometime ago by Lynn~\cite{lynn}. This
description has recently been given a more realistic structure by
Lutz, Friman and Appel~\cite{lutz}. The idea here is that one
writes down an {\it effective potential or energy} calculated to
the highest order feasible in practice in chiral perturbation
theory,
suitably taking into account {\it all} relevant scales involved
and then looks for the minimum of the effective potential to be
identified with the nuclear matter ground state. The state so
obtained may be identified with Lynn's chiral liquid state. The
connection between the skyrmion with an infinite winding number
and the chiral liquid matter -- which must exist in large $N_c$
limit -- is presently not understood.

The starting point of our consideration is the assumption that we
have a chiral-liquid solution of the type described in \cite{lutz}
that represents the ground state (``vacuum"), on top of which
fluctuations can be calculated. The discussion of \cite{lutz} does
not specify how these fluctuations are to be made. To proceed, we
propose that the parameters of the Lagrangian (such as masses,
coupling constants etc) of the fields representing the relevant
degrees of freedom are determined at this ground state, not at the
zero-density vacuum which gives the starting point of
the Lynn strategy and hence run with density~\footnote{The meaning
of density-dependence in the parameters in an effective Lagrangian
we shall study will be precisely defined later. There is a
subtlety due to the requirement of chiral symmetry that needs to
be addressed.}. The Lagrangian so defined is assumed to satisfy
the same symmetry constraints -- such as chiral symmetry and scale
anomaly -- as those intrinsic to QCD at zero density.
\subsection{Effective chiral Lagrangian}
\itt
 Let us denote the parameters so defined at a density $\rho$
with a star. The mass of a nucleon in the system will be denoted
as $M^\star$, the pion decay constant $f_\pi^\star$ etc. The
simplest chiral Lagrangian for the nuclear system so defined takes
the form
\be
\calL=\bar{N}[i\gamma_{\mu}(\del^{\mu}+iv^{\mu} +g_A^\star\gamma_5
a^{\mu}) -M^\star]N -\sum_i C_i^\star (\bar{N}\Gamma_i N)^2 +\cdots
\label{leff}
\ee
where the ellipsis denotes higher dimensional nucleon operators and
the $\Gamma_i$'s Dirac and flavor matrices as well as derivatives
consistent with chiral symmetry. Furthermore
 \be
\xi^2 &=& U =e^{i{\bf \pi}\cdot{\bf \tau} /f_\pi^\star}\nonumber\\
v_\mu &=& -\frac{i}{2}(\xi^{\dagger}\del_\mu\xi
+\xi\del_\mu\xi^\dagger )\nonumber\\ a_\mu &=&
-\frac{i}{2}(\xi^\dagger\del_\mu\xi -\xi\del_\mu\xi^\dagger ).
\ee
In (\ref{leff}) only the pion ($\pi$) and nucleon ($N$) fields
appear explicitly; all other fields have been integrated out. The
effect of massive degrees of freedom will be lodged in
higher-dimensional and/or higher-derivative interactions. The
external electro-weak fields that we will consider below are
straightforwardly incorporated by suitable gauging.

If one is considering symmetric nuclear matter and limits oneself
to the mean field approximation, one can write, following
\cite{gelmini}, an equivalent Lagrangian that contains just the
degrees of freedom that figure in a linear model of the
Walecka-type ~\cite{waleckamodel}
\be
\calL &=& \bar{N}(i\gamma_{\mu}(\del^\mu+ig_v^\star\omega^\mu
)-M^\star+h^\star\phi )N \nonumber\\ & &-\frac 14 F_{\mu\nu}^2
+\frac 12 (\partial_\mu \phi)^2
+\frac{{m^\star_\omega}^2}{2}\omega^2
-\frac{{m^\star_\phi}^2}{2}\phi^2+\cdots\label{leff2} \ee
 where the ellipsis denotes higher-dimension operators.
 This Lagrangian is totally equivalent to (\ref{leff}) in the mean
field approximation~\cite{gelmini,BR96}.
Unless otherwise noted, we will be using (\ref{leff}).
\subsection{Interpreting the density dependence of the parameters}
\itt
 {}From a field theory point of view, it is unclear what
``density dependence" of various constants of the Lagrangian
means. This is because the number density $\rho$ is defined as the
expectation value of the number density operator $\bar{N}\gamma_0
N$ with respect to the state vector one is considering. Thus the
density $\rho$ is defined only once the state is determined. The
only way that such a quantity can be introduced into the
Lagrangian is to assume that the parameters of the Lagrangian such
as coupling constants and masses are functions of the fields
involved. The constraint that the Lagrangian be invariant under
chiral symmetry transformation then limits the field dependence.
One may choose a chiral singlet scalar field or chiral invariant
bilinear in the nucleon fields. In what follows we shall choose
the latter.

For this purpose, we define the chirally invariant operator
\be
\check\rho u^\mu \equiv \bar{N}\gamma^\mu N
\ee
where
\be
u^\mu =\frac{1}{\sqrt{1-\vb^2}}(1,\vb )
=\frac{1}{\sqrt{\rho^2-\jb^2}}(\rho ,\jb ).
\ee
is the fluid 4-velocity. Here
\be
\jb = \langle \bar{N}\gammab N \rangle \ee is the baryon current
density and
\be
\rho =\langle N^\dagger N \rangle =\sum_in_i \label{rhon} \ee the
baryon number density. The expectation value of $\check\rho$
yields the baryon density in the rest-frame of the fluid. Using
$\check\rho$ it is easy to construct a Lorentz invariant, chirally
invariant Lagrangian with density dependent parameters. However,
here we shall not use the relativistic formulation.

Now a density dependent mass parameter in the Lagrangian should be
interpreted as
\be
m^\star=m^\star (\check\rho).
\ee
This means that the model (\ref{leff2}) is no longer linear, but
highly non-linear even at the mean field level. We shall illustrate
this using the Lagrangian (\ref{leff2}) in the mean field
approximation and show that our interpretation is thermodynamically
consistent.

The Euler-Lagrange equations of motion for the bosonic fields are
the usual ones but the nucleon equation of motion is not. This is
because of the functional dependence of the masses and coupling
constants on the nucleon field:
\be
\frac{\delta\calL}{\delta\bar{N}}&=&
\frac{\del\calL}{\del\bar{N}}+\frac{\del\calL}{\del\check\rho}\frac{\del\check\rho}
{\del\bar{N}} \nonumber\\ &=&[i\gamma^\mu (\del_\mu
+ig_v^\star\omega_\mu -iu_\mu\check\Sigma ) -M^\star +h^\star\phi
] N\nonumber\\ &=&0\label{fer} \ee with
\be
\check\Sigma &=&\frac{\del\calL}{\del\check\rho}\\ &=&
m_\omega^\star\omega^2\frac{\del m_\omega^\star}{\del\check\rho}
-m_\phi^\star\phi^2\frac{\del m_\phi^\star}{\del\check\rho}
-\bar{N}\omega^\mu\gamma_\mu N\frac{\del
g_v^\star}{\del\check\rho} -\bar{N}N\frac{\del
M^\star}{\del\check\rho}. \nonumber
\ee
Here we are assuming that $(\partial/\partial\check\rho)
h^\star\approx 0$. It may be possible to justify this but we shall
not attempt it here. The additional term $\check\Sigma$, which may
be related to what is referred to in many-body theory as
``rearrangement terms", is essential in making the theory
consistent. This point has been overlooked in the literature.

Here we shall briefly summarize the results. Details can be found
in \cite{consistency,songPR}. When one computes the energy-momentum
tensor with (\ref{leff2}), one finds the canonical term, which is
obtained when the parameters are treated as constants, as well as a
new term proportional to $\check\Sigma$
\be
T^{\mu\nu}=T_{can}^{\mu\nu}+\check\Sigma(\bar{N}{u\cdot\gamma} N)
g^{\mu\nu}.\label{tmunu}
\ee
The pressure
is then given by $\frac13
\langle T_{ii}\rangle_{\vb =0}$. The additional term in (\ref{tmunu})
matches precisely the terms that arise when the derivative with
respect to $\rho$ acts on the density-dependent masses and coupling
constants in the formula derived from $T_{00}$:
\be
p=-\frac{\del E}{\del V}=\rho^2\frac{\del\E
/\rho}{\del\rho}=\mu\rho -\E \ee
 where
\be
\E=\langle T^{00}\rangle.
\ee
This matching assures energy-momentum conservation and
thermodynamic consistency.

Once the interpretation of the density dependence is specified, the
derivation of the  Landau-Migdal parameters, thermodynamic
quantities etc. from (\ref{leff2}) is completely analogous to the
procedure used by Matsui~\cite{matsui} for Walecka's linear
$\sigma-\omega$ model.
\subsection{Nuclear matter with BR scaling}
\itt
We saw above that the masses and coupling constants in
(\ref{leff2}) (or equivalently (\ref{leff})) are to be treated as
functionals of $\check\rho$ where
\be
\check\rho u^\mu \equiv \bar{N}\gamma^\mu N.
\ee
When treated at the mean field level, $\check\rho$ is just the
number density, so the parameters become density-dependent. The
dependence of the parameters in the Lagrangian on the fields rather
than on the density is essential for thermodynamic consistency.
Note however that these considerations do not require the
parameters to satisfy scaling relations. It is the chiral symmetry
and scale symmetry that suggest that the masses satisfy BR scaling
at the mean field level~\cite{BR91}
\be
\Phi (\rho) \approx \frac{f^\star_\pi (\rho )}{f_\pi} \approx
\frac{m^\star_\phi (\rho )}{m_\phi}\approx\frac{m_V^\star (\rho
)}{m_V} \approx \frac{M^\star (\rho )}{M}.
\label{BRscaling}
\ee
Here $V$ stands for the light-quark vector mesons $\rho$ and
$\omega$. The quantity $\Phi(\rho)$ is the scaling factor that
needs to be determined from theory or experiments. For
concreteness, we shall assume
\be
\Phi (\rho)=(1+y\rho/\rho_0)^{-1}.\label{y}
\ee
The value of $y$ will be determined below by a global fit of the
ground state properties of nuclear matter. Now taking the
free-space values,
\be
M=938, \ \ m_\omega=783, \ \ m_\phi=700\ \
{\rm MeV}
\ee
and
\be g_v=15.8, \ \ h=6.62,
\ee
with one additional assumption that the vector coupling $g_v^\star$
scales like the mass $m_\omega^\star$ and $h^\star$ is almost
constant, one finds the following properties for the ground state
of nuclear matter
\be
m_N^\star/M=0.62, \ \ E/A-M=-16.0\ \ {\rm MeV}, \ \ k_F=257\ \ {\rm
MeV}, \ \ K=296\ \ {\rm MeV}.
\label{fit}
\ee
Here $k_F$ is the Fermi momentum at the saturation point and $K$ is
the corresponding compression modulus. The best values favored by
Nature that are ``well determined" and that ``can be associated
with an equal number of nuclear properties and general features of
RMF (relativistic mean field) models"~\cite{bestvalues}\footnote{It
is worth pointing out that the RMF that has been successful so far
involves non-linear terms deemed ``natural" in the terminology of
EFT. These terms can be interpreted as representing high-dimension
Fermi operators.} are
\be
m_N^\star/M=0.61\pm 0.03, \ \ E/A-M=-16.0\pm 0.1\ \ {\rm MeV},
\nonumber \\ k_F=256\pm 2 \
\
{\rm MeV}, \ \ K=250\pm 50 \ \ {\rm MeV}.
\ee
To arrive at (\ref{fit}), we need $y=0.28$ which implies that $\Phi
(\rho_0)\approx 0.78$. The scaling of $g_v$, which is needed to
obtain a good fit, was not incorporated in the original BR
scaling~\cite{BR91} but it does not invalidate the scaling relation
(\ref{BRscaling}) which is a mean field relation. The scaling of
the coupling constant is a fluctuation effect on top of the BR
scaling ground state, that is, a running as in the renormalization
group as discussed in \cite{kim}. A caveat here is that at this
level, the KSRF relation that holds in free space between the
vector mass $m_V$ and $f_\pi g_v$ must have a density-dependent
correction in order for the scaling of $g_v^\star$ to make sense.
To date the possible validity of the KSRF relation or some
generalization of it in medium is not yet unraveled.

Another observation of interest is the in-medium mass of the
scalar $\phi$.~\footnote{The scalar that figures here is an
effective degree of freedom which need not be identified with a
particle in the Particle Data booklet. From our point of view, it
is closer to the ``dilaton" discussed by Beane and van
Kolck~\cite{beane}. } In the analysis of \cite{bestvalues}, the
scalar mass does not have a simple scaling since it is a
complicated non-linear theory. See \cite{songPR} for a detailed
discussion on this matter. In the present theory, we in fact have
the relation
\be
m_\phi^\star (\rho_0)=m_\phi\Phi (\rho_0)
\ee
which for $\Phi (\rho_0)\approx 0.78$ gives the mass of the scalar
in nuclear matter to be 546 MeV which should be compared with the
value $500\pm 20$ MeV favored by {\cite{bestvalues}.

It should be stressed that given the simplicity of the model
considered here, the agreement between the simple BR scaling model
and the sophisticated non-linear mean-field model~\cite{bestvalues}
is most remarkable. Whether there is something deep here or it is
just a coincidence is an issue to be resolved.

\section{Deriving Migdal's Formulas from Effective Chiral
Lagrangians}
\itt
Here we sketch Migdal's derivation of nuclear orbital gyromagnetic
ratio and then write an analogous expression for the nuclear axial
charge operator following the same steps taken for the vector
current. We have no rigorous proof that the axial charge that
results is a unique one that follows from the premise of Fermi
liquid theory but we are offering it here as a {\it possible} one.
\subsection{Landau-Migdal formulation}
\subsubsection{Vector currents}
\itt
 Consider the response of a heavy nucleus to a slowly varying
electromagnetic field. We wish to calculate the gyromagnetic ratio
$g_l$ of a nucleon sitting on top of the Fermi sea. There are
several ways for doing this calculation~\cite{FRS99}. Here we shall
use the simplest which turns out to be straightforwardly applicable
to the axial current, in particular to its time component, i.e.,
the axial charge.

We are interested in the response of a {\it homogeneous}
quasiparticle excitation to the convection current. This
corresponds to the limit $q/\omega\rightarrow 0$ where
$(\omega,{\Bq})$ is the four-momentum transfer induced by the
electromagnetic field. The first step is to compute the total
current carried by the wave packet of a {\it localized}
quasiparticle with group velocity ${\vb}_F=\frac{\Bk}{m_N^\star}$
where $m_N^\star$ is the Landau effective mass of the
quasiparticle and ${\Bk}$ is the momentum carried by the
quasiparticle~\footnote{This should be distinguished from the BR
scaling effective mass $M^\star$ that appears in (\ref{leff}) and
(\ref{leff2}) and will be defined more precisely later.}. The
convection current for a {\it localized quasiparticle} is \be
\Jb_{LQP}=\frac{\Bk}{m_N^\star} \left(\frac{1+\tau_3}{2}\right)
\label{locQP}. \ee However this is not really what we want. Among
other things, it does not conserve the charge. This is because
the quasiparticle interacts with the surrounding medium
generating what is known as ``back-flow." Consequently we have to
incorporate the back-flow to restore gauge invariance. A simple
way to account for the back-flow is to compute the particle-hole
contributions of the type given in Fig.\ref{ph} with the full
particle-hole interaction -- represented in the figure by the
solid circle given by Eq.~(\ref{qpint}) -- in the limit that
$\omega/q\rightarrow 0$. (Note that this contribution vanishes in
the other limit $q/\omega\rightarrow 0$.)
\begin{figure}
\centerline{\epsfig{file=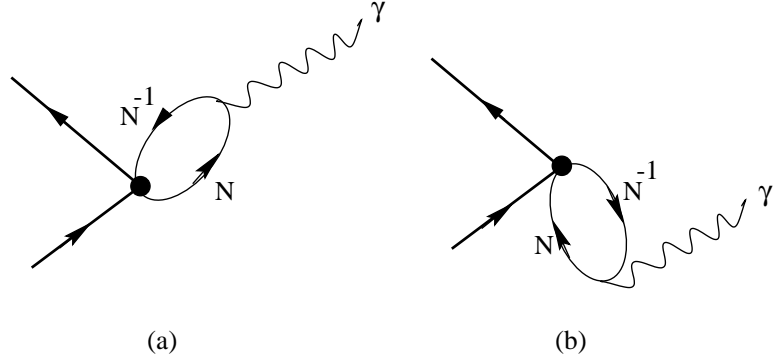,width=4in}} \caption{\small
Particle-hole contributions to the convection current involving
the full particle-hole interaction (solid circle) given by
Eq.~(\ref{qpint}). Here the backward-going nucleon line $N^{-1}$
denotes a hole and the wiggly line the photon. These graphs
vanish in the $q/\omega\rightarrow 0$ limit.}\label{ph}
\end{figure}
The full interaction between two quasiparticles ${\pb_1}$ and
${\pb_2}$ at the Fermi surface of symmetric nuclear matter written
in terms of a few spin and isospin invariants is \cite{BSJ}
\be
\label{qpint}
{ f}_{\pb_1\sigma_1\tau_1,\pb_2\sigma_2\tau_2}&=&
\frac{1}{N(0)}\left[F(\cos \theta_{12})+F^\prime(\cos
\theta_{12})\taub_1\cdot \taub_2+G(\cos \theta_{12})\sigmab_1\cdot
\sigmab_2\phantom{\frac{\qb^{\,2}}{k_F^2}}\right.\nonumber\\
&+&\left. G^\prime(\cos
\theta_{12})\sigmab_1\cdot \sigmab_2\taub_1\cdot\taub_2
+\frac{\qb^{\, 2}}{k_F^2}H(\cos
\theta_{12})S_{12}(\qbhat)\right.\nonumber\\
&+&\left.\frac{\qb^{\, 2}}{k_F^2}H^\prime(\cos \theta_{12})S_{12}
(\qbhat)\taub_1\cdot\taub_2\right]
\ee
where $\theta_{12}$ is the angle between ${\pb_1}$ and ${\pb_2}$
and $N(0)=\frac{\gamma
k_F^2}{(2\pi^2)}\left(\frac{dp}{d\ve}\right)_F$ is the density of
states at the Fermi surface. We use natural units with $\hbar=1$.
The spin and isospin degeneracy factor $\gamma$ equals 4 in
symmetric nuclear matter. Furthermore, $\qb=\pb_1-\pb_2$ and
\bq
\label{tensor}
S_{12}(\qbhat) = 3\sigmab_1\cdot\qbhat\sigmab_2\cdot\qbhat -
\sigmab_1\cdot\sigmab_2,
\eq
where $\qbhat = \qb/|\qb |$. The
tensor interactions $H$ and $H^\prime$ are important for the axial
charge which we will consider later. The functions $F, F^\prime,
\dots$ are expanded in Legendre polynomials,
\bq
F(\cos \theta_{12})=\sum_\ell F_\ell
P_\ell(\cos \theta_{12}),
\eq
with analogous expansion for the spin- and isospin-dependent
interactions.

In terms of (\ref{qpint}), the quasiparticle-quasihole graphs of
Fig.\ref{ph}, suitably generalized to the full interaction, yield
\be
\Jb_{ph} &=&-\frac{1}{3\pi^2}\kbhat k_F^2(f_1+f_1^\prime
\tau_3)\nonumber\\ &=&-\frac{\Bk}{M}\left( \frac{\tilde{F}_1
+\tilde{F}_1^\prime\tau_3}{6}\right)\label{parthole}
\ee
where $M$ denotes the free-space mass of the nucleon and
\be
\tilde{F_i}\equiv (M/m_N^\star) F_i.
\ee
In order to obtain the desired current, we have to add the backflow
term (i.e., $-\Jb_{ph}$) to the localized quasiparticle term
(\ref{locQP}),
\be
\Jb_{migdal}=\Jb_{LQP}-\Jb_{ph}=\frac{\Bk}{M}g_l =
\frac{\Bk}{M} \left(\frac{1+\tau_3}{2}+ \frac{1}{6}
(\tilde{F}^\prime_1-\tilde{F}_1)\tau_3\right),\label{Jtotal}
\ee
where
\bq
g_l=\frac{1+\tau_3}{2}+\delta g_l\label{gl}
\eq
is the orbital gyromagnetic ratio and
\bq
\delta g_l=\frac{1}{6}
(\tilde{F}^\prime_1-\tilde{F}_1)\tau_3.\label{deltagyro}
\eq
In arriving at (\ref{Jtotal}), we have used the relation between
the Landau effective mass and the quasiparticle interaction
\be
\frac{m_N^\star}{M}=1+\frac 13 F_1=(1-\frac 13
\tilde{F}_1)^{-1}.\label{landaumass}
\ee

It is important to note that, as a consequence of charge
conservation and Galilei invariance the isoscalar term in
(\ref{Jtotal}) is not renormalized by the interaction. Thus, the
renormalization of $g_l$ is purely isovector. It is also important
to note that it is the free-space mass $M$, not the Landau mass
$m_N^\star$, that appears in (\ref{Jtotal}). This is an analog to
Kohn's theorem for the cyclotron frequency of an electron in an
external magnetic field~\cite{kohn,hkm98}~\footnote{The cyclotron
frequency of an electron with a Landau effective mass $m_e^\star$
in an external magnetic field of magnitude $B=2\pi n_f\phi/e$
where $n_f$ is the electron number density and $\phi$ is the flux
integer (=2 for fermions) is {\it not} $\omega_0=2\pi
n_f\phi/m_e^\star$ as one would naively expect for a localized
quasiparticle but $\omega_0=2\pi n_f\phi/m_e$ due to the
back-flow effect.}, and constitutes a strong constraint for a
consistent theory to satisfy. The effective Lagrangian theory
discussed below does satisfy this condition.

\subsubsection{Axial currents}
\itt
Next we turn to the axial charge operator $A_0^a$ (where the
superscript $a$ is an isospin index). In deriving the ``Migdal
formula" for this operator~\footnote{We put quotation marks since
Migdal did not derive formulas for the axial charge.}, we assume
that we can follow the exactly the same reasoning as above for the
vector current. This assumption needs still to be justified.

In matter-free space,  the axial charge operator for a
non-relativistic nucleon with mass $M$ is
\be
A_0^a=g_A\frac{\tau^a}{2} \frac{\Bsigma\cdot\Bk}{M},
\ee
while in dense matter a {\it localized quasiparticle} with a Landau
effective mass $m_N^\star$ has the axial charge
\be
A_{0LQP}^a=g_A\frac{\tau^a}{2} \frac{\Bsigma\cdot\Bk}{m_N^\star}.
\ee
Next we calculate the particle-hole contribution -- which is minus
the back-flow contribution -- in the same way as for the vector
current (i.e., Fig.\ref{ph} with the pion exchange replaced by $f$,
eq.(\ref{qpint})). The result~\cite{FRS99} is
\be
A_{0ph}^a=-g_A\frac{\tau^a}{2}
\frac{\Bsigma\cdot\Bk}{m_N^\star}\Delta
\ee
with \be \Delta=\frac 13 G_1^\prime -\frac{10}{3}H_0^\prime+\frac
43 H_1^\prime -\frac{2}{15}H_2^\prime \ee where $G^\prime$ and
$H^\prime$ are the spin-isospin-dependent components of the force
given in Eq.(\ref{qpint}). Therefore the total is
\be
A_{0migdal}^a=A_{0LQP}^a - A_{0ph}^a=g_A\frac{\tau^a}{2}
\frac{\Bsigma\cdot\Bk}{m_N^\star}(1+\Delta).\label{a0migdal} \ee
It will become clearer when we calculate the same quantity based
on chiral Lagragian, but at this point it should be noted, that
unlike the vector current, here there is no analog of the Kohn
theorem. This is because chiral symmetry is realized, not in the
Wigner mode, but rather in the Goldstone mode for which the
meaning of a conserved charge is different from that of the vector
charge.
 Another
point to be noted is that while for the convection current, only
$F_1$ and $F_1^\prime$ appear, it is a lot more complicated for the
axial charge. It involves spin-isospin dependent interactions as
well as tensor forces. These two features will show up
non-trivially when we compute the $\delta g_l$ and the $\Delta$
with the effective chiral Lagrangian.

\subsection{Calculation with effective chiral Lagrangian}
\itt
We will now compute $\delta g_l$ and $\Delta$ using a BR scaling
chiral Lagrangian. One can use either the Lagrangian (\ref{leff})
or the Lagrangian (\ref{leff2}) with BR scaling~\cite{BR91}
incorporated. We shall use (\ref{leff}) as we did for the vector
current. We need only the two terms of the four-Fermi interactions
that correspond to the $\omega$ and $\rho$ channels:
\be
\label{four-Fermion} -\frac{{C_\omega^\star}^2}{2}
(\bar{N}\gamma_\mu N)^2
-\frac{{C_\rho^\star}^2}{2}(\bar{N}\gamma_\mu\tau N)^2 +\cdots,
\ee i.e., what remains when the vectors $\omega$ and $\rho$ are
integrated out. The subscripts represent not only the vector
mesons $\omega$ and $\rho$ nuclear physicists are familiar with
but also {\it all} vector mesons of the same quantum numbers, so
the two ``counter terms" subsume the {\it full} short-distance
physics of the same chiral order.

\subsubsection{Landau mass from chiral Lagrangian}
\itt
We first calculate the single-particle energy with (\ref{leff}). In
the nonrelativistic approximation, we have
\be
\ve_p =\frac{p^2}{2 M^\star}+{C_\omega^\star}^2\langle N^\dagger
N\rangle +\Sigma_\pi (p),\label{epsilon}
\ee
where $M^\star = \Phi M$ is the BR-scaling nucleon mass and
$\Sigma_\pi (p)$ is the nucleon self-energy due to the pion Fock
term. The Landau effective mass is defined~\cite{FR96} by
\be
\frac{m_N^\star}{M}=\frac{k_F}{M} \left( \left.
\frac{d}{dp}\ve_p\right|_{p=k_F}\right)^{-1} =\left(\Phi^{-1}
-\frac 13 \tilde{F}_1 (\pi)\right)^{-1}\label{mstar2}
\ee
where we have used the fact that the second term of (\ref{epsilon})
does not contribute and
\be
\tilde{F}_1(\pi )=-\frac{3M}{k_F} \left. \frac{d\Sigma_\pi
(p)}{dp}\right|_{p=k_F} = -\frac{3f^2 M}{8\pi^2 k_F} I_1
\label{piF1}
\ee
where $f=g_Am_\pi/(2f_\pi)\simeq 1$ and
\be
I_1=\int_{-1}^1 dx \frac{x}{1-x+\frac{m_\pi^2}{2k_F^2}}= -2 +
(1+\frac{m_\pi^2}{2k_F^2})\ln
(1+\frac{4k_F^2}{m_\pi^2}).\label{I1}
\ee
Now using the Landau mass formula (\ref{landaumass}) and
\be
\tilde{F}_1=\tilde{F}_1 (\omega)+\tilde{F}_1 (\pi)
\ee
we find
\be
\tilde{F}_1 (\omega)=3(1-\Phi^{-1}).\label{mainrelation} \ee

\subsubsection{Convection current}
\itt
\begin{figure}
\centerline{\epsfig{file=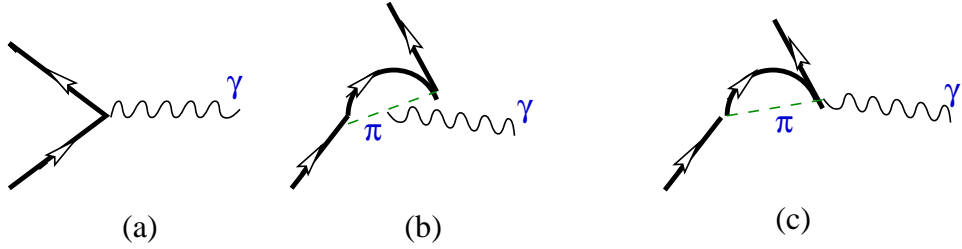,width=5in}} \caption{\small
\label{miyazawa}Feynman diagrams contributing to the EM convection
current in effective chiral Lagrangian field theory. Figure (a) is
the single-particle term and (b, c) the next-to-leading chiral
order pion-exchange current term. Figure (c) does not contribute
to the convection current; it renormalizes the spin gyromagnetic
ratio.}
\end{figure}
In the chiral Lagrangian approach, the isovector magnetic multipole
operator to which the convection current belongs is
chiral-filter-protected~\cite{KDR} which means that the
one-soft-pion exchange should dominate in the correction to the
leading single-particle term. The single-particle term for a
nucleon with the BR scaling mass $M^\star$ on the Fermi surface
with momentum $\Bk$ corresponding to Figure \ref{miyazawa}a is
\be
\Jb_{1-body}=\frac{\Bk}{M^\star} \frac{1+\tau_3}{2}.\label{J1} \ee
Note that the nucleon mass appearing in (\ref{J1}) is the BR
scaling mass $M^\star$ as it appears in the Lagrangian, not the
(Landau) effective mass $m_L^\star$ that enters in the Fermi-liquid
approach for the localized quasiparticle current. To the
next-to-leading order, we have two soft-pion terms
Fig.\ref{miyazawa}a,b as discussed in \cite{KDR}. To the convection
current, only Fig.\ \ref{miyazawa}b contributes
\be
\Jb_{2-body}^\pi=\frac{\Bk}{k_F}\frac{f^2}{4\pi^2} I_1\tau_3 =
\frac{\Bk}{M}\frac{1}{6}
(\tilde{F}^\prime_1(\pi)-\tilde{F}_1(\pi))\tau_3. \label{J2} \ee
 In arriving at this formula, it has been assumed that pion
properties are unchanged in medium up to nuclear matter density.
Since pions are almost Goldstone bosons, this assumption seems
reasonable. Indeed it is consistent with what is observed in
Nature. Note that there are no unknown parameters associated with
the pion contribution (\ref{J2}): the one-pion-exchange
contributions to the Landau parameters $\tilde{F}_1(\pi)$ and
$\tilde{F}_1^\prime(\pi)$ are entirely fixed by the chiral
effective Lagrangian at any density.

The contributions from the four-Fermi interactions (that is, the
vector meson degrees of freedom) are subleading to the pion
exchange by the chiral filter~\cite{KDR}. They are given by Fig.\
\ref{nbar}.
\begin{figure}
\centerline{\epsfig{file=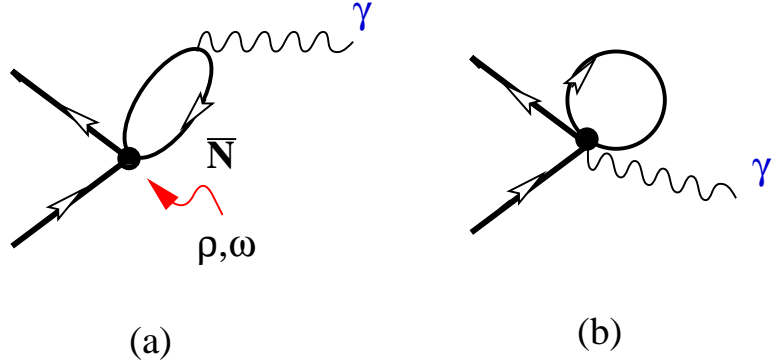,width=4in}} \caption{\small (a)
Feynman diagram contributing to the EM convection current from
four-Fermi interactions corresponding to {\it all channels} of the
$\omega$ and $\rho$ quantum numbers (contact interaction indicated
by the blob) in effective chiral Lagrangian field theory. The
$\bar{N}$ denotes the anti-nucleon state that is given in the
chiral Lagrangian as a $1/M$ correction and the one without arrow
is a Pauli-blocked or occupied state. (b) The equivalent graph in
heavy-fermion formalism with the anti-nucleon line shrunk to a
point. The blob represents a four-Fermi interaction coupled to a
photon that enters in (\ref{leff}) as a $1/M$ counter term.
}\label{nbar}
\end{figure}
Both the $\omega$ (isoscalar) and $\rho$ (isovector) channels
contribute through the antiparticle intermediate state as shown in
Fig.\ \ref{nbar}a. The antiparticle is explicitly indicated in the
figure. However in the heavy-fermion formalism, the backward-going
antinucleon line should be shrunk to a point as Fig.\ \ref{nbar}b,
leaving behind an explicit $1/M^\star$ dependence folded with a
factor of nuclear density indicating a $1/M^\star$ correction in
the chiral expansion~\footnote{The heavy-baryon formalism must be
unreliable once the $M^\star$ drops for $\rho\gsim \rho_0$. One
would then have to resort to a relativistic
formulation~\cite{becher}. We expect, however, that our reasoning
will remain qualitatively intact.}. One can interpret Fig.\
\ref{nbar}a as saturating the corresponding counter term although
this has yet to be verified by writing down the full set of counter
terms at the same order.  We find
\be
\Jb_{2-body}^{\omega}&=&\frac{\Bk}{M}\frac 16 \tilde{F}_1
(\omega), \label{omega}\\ \Jb_{2-body}^{\rho}&=&\frac{\Bk}{M}\frac
16 \tilde{F}_1^\prime (\rho)\tau_3,\label{rho} \ee
 where
\bq\label{f1-omega}
 F_1(\omega)=
-{C_\omega^\star}^2\frac{2k_F^3}{\pi^2M^\star} \eq
 and
 \bq\label{f1p-rho}
F_1^\prime(\rho)= -{C_\rho^\star}^2\frac{2k_F^3}{\pi^2M^\star}.
\eq

The total current given by the sum of (\ref{J1}), (\ref{J2}),
(\ref{omega}) and (\ref{rho}) precisely agrees with the
Fermi-liquid theory result (\ref{Jtotal}) when we identify
\be
\tilde{F}_1&=&\tilde{F}_1 (\omega)+\tilde{F}_1 (\pi),\\
\tilde{F}_1^\prime &=&\tilde{F}_1^\prime (\rho)+\tilde{F}_1^\prime
(\pi). \ee
 If we further assume the same flavor symmetry as in free space
 holds in medium, then
 \be
 \tilde{F}^\prime (\rho)=\tilde{F}^\prime (\omega)/9
 \ee
 which uses the nonet symmetry and
 \be
\tilde{F}^\prime (\pi)=-\tilde{F}(\pi)/3 \ee
 which uses the isotopic invariance.
 The BR scaling chiral Lagrangian prediction reduces to
a one-parameter formula
\be
\delta g_l=\frac 16 (\tilde{F}_1^\prime -\tilde{F}_1)\tau_3 =\frac
49 \left[\Phi^{-1} -1- \frac 12 \tilde{F}_1 (\pi)\right]\tau_3.
\label{chptd} \ee
 Here $\Phi (\rho)$ is the only parameter in the theory that needs
to be determined from theory or experiment. As mentioned,
$\tilde{F}_1 (\pi)$ is fixed for an arbitrary density from the
(assumed) chiral symmetry. It is important that the result is
consistent with charge conservation and Galilei invariance

\subsubsection{Axial charge}
\itt
The axial charge operator in nuclear matter is protected by the
chiral filter in the chiral Lagrangian formalism, so all we need is
the soft-pion exchange implemented with BR scaling. We shall
continue assuming that pions do not scale in medium. It has been
shown in \cite{PMR} that higher order chiral corrections -- such as
loops, higher derivative and four-Fermi terms
 -- to the soft-pion
contribution are small. This means that we can limit ourselves to
the tree order in the chiral counting and to the pionic range with
shorter-range interactions subsumed in the BR scaling.

The procedure for the case at hand will then be identical to that
we used for the convection current. The axial charge for a single
particle will be identical to that of a particle in free space
except that the nucleon mass $M$ is to be replaced by the BR
scaling mass $M^\star$
\be
A_0^a=g_A\frac{\tau^a}{2} \frac{\Bsigma\cdot\Bk}{M^\star}.
 \ee
 Now the leading correction to the single-particle term is given
by a diagram similar to Fig.\ref{miyazawa}c with the photon
replaced by the weak axial charge. There is no term equivalent to
Fig.\ref{miyazawa}b due to G-parity invariance. The calculation is
straightforward and the result is
\be
{\A0}^i_{2-body}=g_A\frac{\Bsigma\cdot\Bk}{M^\star}\frac{\tau^i}{2}
\tilde{\Delta} \label{a02bod} \ee with
\be
\tilde{\Delta}=\frac{f^2 k_F M}{2g_A^2m_\pi^2\pi^2}\left(I_0-I_1-
\frac{m_\pi^2}{2k_F^2}I_1\right)\label{Delta} \ee
 where $I_1$ is as defined in (\ref{I1}) and $I_0$ is
 \be
I_0=\int_{-1}^1 dx\frac{1}{1-x+\frac{m_\pi^2}{2k_F^2}}=
\ln\left(1+\frac{4k_F^2}{m_\pi^2}\right).
 \ee
 The factor $(1/g_A^2)$ in (\ref{Delta}) arose from replacing
$\frac{1}{f_\pi^2}$ by $\frac{g_{\pi NN}^2}{g_A^2 M^2}$ using the
free-space Goldberger-Treiman relation.

Collecting all term, the chiral Lagrangian prediction is
\be
{\A0}^a_{chiral}=g_A\frac{\Bsigma\cdot\Bk}{M^\star}\frac{\tau^a}{2}
(1+\tilde{\Delta}).
\label{chiral}
\ee
For comparison with the ``Migdal formula" ${\A0}^i_{migdal}$, we
re-express $1/M^\star$ in terms of $1/m_N^\star$
\be
1/M^\star=\frac{1}{m_N^\star} (1-\frac{\Phi}{3} \tilde{F}_1
(\pi))^{-1}.
\ee
Thus
\be
{\A0}^a_{chiral}=g_A\frac{\Bsigma\cdot\Bk}{m_N^\star}\frac{\tau^a}{2}
(1+\tilde{\Delta}^\prime)
\label{chiral2}
\ee
where
\be
\tilde{\Delta}^\prime=(\tilde{\Delta}+\frac{\Phi}{3} \tilde{F}_1
(\pi)) (1-\frac{\Phi}{3} \tilde{F}_1 (\pi))^{-1}.
\ee
Comparing with the ``Migdal formula" (\ref{a0migdal}), we obtain a
formula that expresses a combination of spin-isospin-dependent
Landau-Migdal parameters in terms of constants that figure in the
chiral Lagrangian with BR scaling:
\be
\frac 13 G_1^\prime -\frac{10}{3}H_0^\prime+\frac 43
H_1^\prime -\frac{2}{15}H_2^\prime=\tilde{\Delta}^\prime.
\label{axialrelation}
\ee
Again the result depends on only one parameter $\Phi$.

There are two points to note here. One is as noted in the
Landau-Migdal formulation that there is no equivalent to ``Kohn's
theorem" for the axial charge. The other is that the soft-pion
contribution combined with BR scaling does not lend itself to a
direct term-by-term identification on both sides. These are all
different from the case of the convection current. In the axial
case, both the Landau-Migdal approach and the chiral Lagrangian
approach involve complicated relations: on the right-hand side of
(\ref{axialrelation}), the factor $g_A$ appears in a non-trivial
way and exhibits features that are characteristic of the
spontaneously broken axial symmetry and on the left-hand side, this
complexity is manifested by the fact that, due to the tensor force,
the Migdal parameters involved comprise several multipoles ($l=0,
1, 2$) of the quasiparticle interaction.

\section{Comparison with Experiments}
\itt
In confronting our theory with Nature, we shall assume that data on
heavy nuclei represent nuclear matter. This aspect has been
extensively discussed elsewhere so we shall be brief.
\subsection{Extracting $\Phi (\rho_0)$}
\itt
If one assumes BR scaling, then there are several ways to determine
$\Phi$ at normal nuclear matter density. We shall mention three of
them.
\begin{enumerate}
\item
The first way is that if pions are taken to be non-scaling, then
the in-medium Gell-Mann-Oakes-Renner relation
\be
m_\pi^2 {f_\pi^\star}^2=-(m_u+m_d) \langle \bar{q}q\rangle^\star
\ee
gives
\be
\frac{f_\pi^\star}{f_\pi}\approx\left(\frac{\langle
\bar{q}q\rangle^\star}{\langle \bar{q}q\rangle_0}\right)^{1/2}.
\ee
>From the value of quark condensate in nuclear matter estimated from
the empirical $\pi N$ sigma term and using Feynman-Hellmann theorem
in the linear density approximation~\footnote{The linear density
approximation may be suspect already at nuclear matter density, so
it is difficult to assess the uncertainty involved with this
estimate.}, one finds~\cite{BR96}
\be
\frac{f_\pi^\star}{f_\pi}\approx 0.78.
\ee
\item
The second evidence comes from the property of nuclear matter in
chiral Lagrangian models with BR scaling. A global fit
yields~\cite{songPR}
\be
M^\star/M\approx 0.78\pm 0.02.
\ee
\item
The third evidence comes from QCD sum rule calculation of the mass
of the vector meson $\rho$ in medium~\cite{hatsuda,jin}. The
result is~\cite{jin}
\be
m_\rho^\star/m_\rho= 0.78\pm 0.08. \ee
\end{enumerate}
All three methods give the same result. We are therefore led to use
\be
\Phi (\rho_0)=0.78.
\ee
As a smooth interpolation which seems reasonable at least up to
$\rho\simeq \rho_0$, we take
\be
\Phi (\rho)=(1+0.28\rho/\rho_0)^{-1}\label{scaling}.
\ee
\subsection{The orbital gyromagnetic ratio}
\itt
Given the scaling factor $\Phi (\rho_0)\approx 0.78$ and the pionic
contribution (\ref{piF1}) which at nuclear matter density yields
$\tilde{F}_1 (\pi)\approx -0.459$, the anomalous orbital
gyromagnetic ratio turns out to be
\be
\delta g_l=\frac 49 \left[\Phi^{-1} -1- \frac 12 \tilde{F}_1
(\pi)\right]\tau_3=0.227\tau_3.
\label{deltaglth}
\ee
This is to be compared~\footnote{The precise agreement is probably
coincidental.} with the experimental value for the proton obtained
from the giant dipole resonance in the Pb region~\cite{nolte}
\be
\delta g_l^p=0.23\pm 0.03.
\ee

It is worth commenting at this point which assumptions enter into
this calculation and what the possible implication might be. Apart
from the BR scaling, we have assumed (1) that pions do not scale,
(2) the nonet symmetry for the vector mesons and (3) the isospin
symmetry for the pions. The first is based on the observation that
the pion is an almost Goldstone boson and a truly Goldstone boson
would preserve its symmetry as density is increased beyond normal
nuclear matter density. This assumption needs to be verified. The
second is hard to check and remains to be verified. The third is
most probably solid. The upshot of the result is that the charge is
conserved, ``Kohn's theorem" is satisfied and the agreement with
experiment essentially confirms in an average sense the BR scaling
for the nucleon mass.

\subsection{Landau mass for the nucleon}
\itt
A quantity closely related to $\delta g_l$ is the Landau effective
mass $m_N^\star$. Given $\Phi$ and $\tilde{F_1}(\pi)$ for
$\rho=\rho_0$, we obtain from Eq.(\ref{landaumass}) that
\be
m_N^\star=\left(1/0.78+0.153\right)^{-1}M \approx 0.70 M.
\ee
There are two sources of information that can be compared with
this prediction. One is theoretical, namely, the QCD sum-rule
result~\cite{FJL}
\be
\left(\frac{m_N^\star
(\rho_0)}{M}\right)_{QCD}=0.69^{+0.14}_{-0.07}. \ee
 The other is an indirect semi-empirical indication coming from
peripheral heavy ion collisions at the BEVALAC and the
SIS~\cite{mosel}
\be
{m_N^\star}^{HI}\simeq 0.68 M.
\ee
The agreement here is essentially a
re-confirmation of the gyromagnetic ratio (\ref{deltaglth}).

\subsection{Axial-charge transitions in heavy nuclei}
\itt
Before confronting the chiral Lagrangian prediction (\ref{chiral})
(with (\ref{Delta})) with experiments, we compare the left-hand
side of (\ref{axialrelation}) (i.e., ``Migdal's axial charge") with
one-pion exchange only with the right-hand side which is given to
next-to-leading order (NLO) chiral perturbation theory with BR
scaling chiral Lagrangian.\footnote{Modulo a correction less than
10\%, this is valid to next-next-to-leading order (NNLO) in chiral
perturbation theory~\cite{PMR}.}
\begin{figure}
\centerline{\epsfig{file=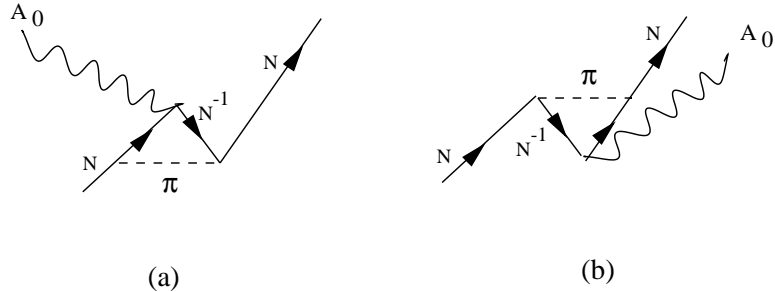,width=4in}} \caption{\small
Particle-hole contribution to the axial charge involving one pion
exchange which is minus the back-flow. Here $A_0$ stands for the
external field probing axial charge, the backward-going nucleon
line $N^{-1}$ denotes a hole and the wiggly line the W boson
connected to the axial charge. These graphs vanish in the
$q/\omega\rightarrow 0$ limit.}\label{ph-pi}
\end{figure}

To compute the Migdal charge, it is sufficient to compute the
one-pion-exchange graphs of Fig.\ref{ph-pi} in the limit
$\omega/q\rightarrow 0$. The negative of this gives the desired
quantity, namely, the ``back-flow." A simple calculation gives
\be (\frac 13 G_1^\prime -\frac{10}{3}H_0^\prime+\frac 43
H_1^\prime -\frac{2}{15}H_2^\prime)_\pi=\frac{f^2k_F
m_N^\star}{4m_\pi^2 \pi^2} (I_0-I_1)\label{Deltapi}
 \ee
 where the subscript $\pi$ denotes the pionic contribution.
 Now since the right-hand side of (\ref{axialrelation}) is valid
 beyond the leading order in chiral perturbation theory, it
contains information that accounts for more than one-pion exchange.
In the same vein, (\ref{Deltapi}), although the interaction is
evaluated with one-pion exchange, contains a lot more since the
mass of the nucleon is given by the Landau mass $m_N^\star$.
Therefore there is no reason to expect a one-to-one correspondence
between the two. Even so, we conjecture that to the extent that the
dynamics is governed by the pion exchange corrected by the BR
scaling $\Phi$, the two must be approximately the same. That is to
say that the combination of the Migdal parameters of
(\ref{axialrelation}) should be saturated by the pions modulo what
corresponds to higher chiral order terms which are argued to be
small. This is required if the chiral filter argument is to hold.

Let us consider how the relation (\ref{axialrelation}) fares with
the pion for $\rho=\rho_0/2$ and $\rho_0$. The left-hand side
-- given by (\ref{Deltapi}) -- comes out to be, respectively,
$0.42$ and $0.50$ for $\rho=\rho_0/2$ and $\rho_0$ while the
right-hand side -- which is the full contribution from the BR
Lagrangian -- gives $0.37$ and $0.55$. Thus the pions are seen to
saturate $\sim 90$ \% of the total predicted by the chiral field
theory with BR scaling.

Although far from direct, there is a beautiful confirmation of the
prediction (\ref{chiral}) from axial charge transitions in heavy
nuclei (denoted by the mass number $A$)
\be
A(J^\pm)\rightarrow A^\prime (J^\mp)+e^-(e^+)+\bar{\nu}(\nu) \ \ \
\Delta T=1. \ee
The quantity we shall look at is Warburton's
$\epsilon_{MEC}$~\cite{warburton} defined by
\be
\epsilon_{MEC}=M_{exp}/M_{sp}
\ee
where $M_{exp}$ is the {\it measured} matrix element for the axial
charge transition and $M_{sp}$ is the {\it theoretical}
single-particle matrix element for a nucleon {\it without} BR
scaling. There are theoretical uncertainties in defining the
latter, so the ratio is not an unambiguous object but what is
significant is Warburton's observation that in heavy nuclei,
$\epsilon_{MEC}$ can be as large as 2:
\be
\epsilon_{MEC}^{HeavyNuclei}=1.9\sim 2.0.
\ee
More recent measurements -- and their analyses -- in different
nuclei~\cite{baumann,minamisono} quantitatively confirm this result
of Warburton.

The theoretical prediction from (\ref{chiral}) is
\be
\epsilon_{MEC}^{chiral}=\Phi^{-1}
(1+\tilde{\Delta})\label{epsilonth}
\ee
with $\tilde{\Delta}$ given by (\ref{Delta}). For nuclear matter
density, we find
\be
\epsilon_{MEC}^{chiral}\approx 2.1.
\ee
The theory therefore describes correctly the large enhancement of
the axial-charge matrix element in nuclei in general and the
density-dependent enhancement in particular. There are two elements
that account for this enhancement. Pions contribute
$\tilde{\Delta}\sim 1/2$ with little density dependence and the BR
scaling $\Phi$ accounts for the further enhancement for heavier
nuclei. This result represents a strong case for the validity of
the theory in the normal density regime.
\section{Going to Denser Matter}
\subsection{Evidence in dense matter?}
\itt The real strength in effective field theories is that one
may be able to describe quantitatively the state of matter that
is formed in high density as one approaches the chiral phase
transition. If one assumes that the matter is a Fermi liquid all
the way to the phase transition, then one can use the BR scaling
chiral Lagrangian in the mean field. But this means that all
degrees of freedom, fermionic as well as bosonic, are treated as
``quasiparticles." It is established that nucleons are
quasiparticles in nuclear matter as Migdal had argued. The shell
model for nuclei is justified by the quasiparticle picture. It is
also supposed that at asymptotic density where weak coupling of
QCD is operative, quarks can be treated as
quasi-quarks~\cite{wilczek}. The presence of a Fermi sea for
nucleons and quarks is one of the ingredients for treating them
as quasiparticles. In the discussions given above, bosons were not
required to be ``quasiparticles" despite that BR scaling is
invoked for both fermions and bosons. In addressing heavy-ion
processes, however, properties of bosons in medium might play an
important role. For instance, in CERN's CERES experiments, it is
the property (i.e., mass, width etc.) of the $\rho$ meson in
dense and hot medium that seems to play a key role. So the
question arises how bosons behave in extreme conditions.

There are some indirect experimental evidences for vector bosons
with dropping masses in dense medium. The effect usually manifests
in spin-isospin dependent nuclear forces and affect spin-isospin
observables~\cite{evidence1,evidence2}. These observables probe
off-shell properties of the mesons involved up to nuclear matter
density and do not in general give direct information on their
``physical" properties in medium. There are similar indications
from tensor forces in heavy nuclei which also can be explained
from the exchange of the $\rho$ meson with a reduced
mass~\cite{BRxxx}. A more direct indication comes from dilepton
production in heavy-ion collisions at CERN CERES. There the
quasi-particle picture of the vector mesons with dropping mass in
hot and dense matter (at a density greater than that of normal
nuclear matter) provides a simple and successful explanation of
the observed downward shift of the invariant mass of the lepton
pair~\cite{LKB}. The approximation used in~\cite{LKB} consists of
taking only tree-order graphs with an effective chiral Lagrangian
a la BR scaling discussed above that gives a realistic description
of nuclear matter: no loop corrections are taken into account in a
proper sense although partial account is of course made in the
unitarization of the amplitudes involved. The question as to what
happens when loop corrections are properly taken into account in
this theory so far remains unanswered. It is also not known
whether the tree-order (i.e., quasi-particle) treatment correctly
describes the excitation of the vector quantum numbers in such
dense matter.

\subsection{Perturbing from zero-density vacuum}
\itt One might attempt an ambitious program to start from an
effective chiral Lagrangian constructed at zero density and do a
systematic chiral expansion to arrive at higher density. This is
the spirit of \cite{lynn,lutz}. Aided by experiments, this
program could be made to work up to nuclear matter density but it
is a completely different matter if one wants to reach a density
where the chiral phase transition can occur. Dense matter probes
short distances and chiral perturbation theory (ChPT) cannot
access such kinematic regime as is clear from Landau-Migdal Fermi
liquid theory. What has been done up to date is a low-order
perturbation calculation in a strong-coupling regime. Now if such
a calculation is based on an effective Lagrangian satisfying
relevant symmetries (e.g., chiral symmetry), leading-order
(tree-order) terms are consistent with low-energy theorems and
should give reasonable results at low density provided the
parameters are picked from experiments. See Rapp and
Wambach~\cite{rapp} for review where the relevant references are
found. In such low-order calculations, one finds that the mesons,
such as the $\rho$ and $a_1$ get ``melted" due to increasing
width and lose their particle identities. However as density
increases away from zero, the tree-order approximations that are
essentially all one can work with cannot be trusted. Exactly
where this discrepancy will become serious is not known. Being in
a strong-coupling regime, anomalous dimensions of certain fields
(such as scalar fields) grow too big to be natural, signaling
that one is fluctuating around the wrong vacuum. We believe this
to be the case already at nuclear matter density. BR scaling
corresponds to shifting to and fluctuating around a ``vacuum"
defined at $\rho\gsim \rho_0$ where the effective coupling gets
weaker in the sense of quasiparticle interactions. As the density
approaches the critical for the chiral phase transition, the
picture with quasi-nucleons goes over to the one with
quasi-quarks. It seems extremely difficult, if not impossible, to
arrive at this picture starting from a strong-coupling hadronic
theory effective at zero density. See \cite{BBR,songPR} for
further discussion on this point.
\subsection{Perturbing from BR scaling ground state}
\itt
 Given a Lagrangian (\ref{leff}) or (\ref{leff2}) with BR scaling
that gives the ground state of nuclear matter correctly, we would
like to know how to make fluctuations around the ground state. As
an illustration, consider kaon-nucleon interactions in
medium~\cite{BR96}. This process is relevant for both laboratory
experiments and for the structure of compact stars as we will
describe below.

For the problem at hand, it is convenient to generalize
(\ref{leff2}) to the $SU(3)$ flavor so as to incorporate kaons in
the Lagrangian. The additional term relevant to the process is
given by \be \delta {\cal
L}_{KN}=\frac{-6i}{8{f^\star}^2}(\overline{N}\gamma_0
N)\overline{K}\del_t K +
\frac{\Sigma_{KN}}{{f^\star}^2}(\overline{N}N)\overline{K}K+\cdots\equiv
{\cal L}_\omega +{\cal L}_\phi+\cdots\label{kaonL} \ee where
$K^T=(K^+ K^0)$, $f^\star$ is the in-medium Goldstone boson decay
constant which within the approximation adopted here, may be
taken to be the pion decay constant and the ellipses stand for
higher-order terms in the chiral counting. The structure of the
first two leading-order terms of the fluctuating Lagrangian is
dictated by current algebras, which implies that $\Sigma_{KN}$ is
the usual KN sigma term in free space and also that it may be
valid near nuclear matter density.

Within the scheme a la BR, we are to work in the mean field
approximation. Assuming that this is valid up to nuclear matter
density, one gets from (\ref{kaonL}) the potential energy for the
scalar ($\phi$) field $S_{K^{-}}$ and the vector ($\omega$) field
$V_{K^{-}}$ that $K^-$ feels in nuclear matter at $\rho=\rho_0$:
\be S_{K^{-}}+V_{K^{-}}\approx -192 \ {\rm MeV}
\label{KNattraction}. \ee For this we have used the value for the
KN sigma term, $\Sigma_{KN}\approx 3.2 m_\pi$ and
$f_\pi^\star/f_\pi\approx\Phi$. The exact value is unknown since
the sigma term is not fixed precisely. The attraction
(\ref{KNattraction}) is consistent with what is observed in
kaonic atoms~\cite{katom} and also with the $K^-/K^+$ production
ratio in heavy-ion collisions at GSI~\cite{LLB}. When applied to
neutron-star matter and extrapolating to higher density, it is
more appropriate to adopt the ``top-down" approach proposed in
\cite{BR96} in which the kaon field is introduced as a matter
field and the relevant fermion field is taken to be the
quasi-quark rather than the nucleon. With a suitable modification
appropriate for the top-down approach of \cite{BR96} in the
Lagrangian (\ref{kaonL}), one then predicts $K^-$ condensation at
a matter density $\rho_c\approx 2\sim 3 \rho_0$ with the
intriguing implication that the maximum stable neutron star mass
is 1.5 times the solar mass~\cite{LLB}. These mean field results
with BR scaling Lagrangians are in agreement with more refined
calculations carried out in high-order chiral perturbation
theory~\cite{CHL}. If it turned out that condensation occurs at
higher density than the range considered so far (due to some
higher order effects that cannot be accessed by the effective
Lagrangian method used), then the presently available machinery
for handling short-distance physics would not be powerful enough
to allow us to pin down the critical density~\cite{pandhari}.
More work is needed in this area.

\subsection{``Sobar" model}
\itt
 Among Migdal's other major contributions to nuclear physics is his
work on pion-nuclear interactions, in particular on pion
condensation in dense nuclear matter~\cite{migdalpi}. It is
suggested that the Fermi liquid description a la BR scaling chiral
Lagrangian can be phrased in a form resembling Migdal's description
of pion condensation. The initial idea is formulated in a series of
recent papers by Kim et al~\cite{sobar}.

\begin{figure} \centerline{\epsfig{file=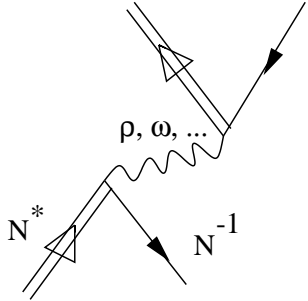,width=4cm}}
\caption{\small Particle-hole coherent modes excited by coupling
to vector mesons $\rho$, $\omega$ ...}\label{coupling}
\end{figure}

Consider a vector meson, say $\omega$, which is inserted in a dense
medium and look at the excitation of {\it coherent} modes of the
$\omega$ quantum number. The $\omega$ meson will be coupled to
particle-hole excitations of the same quantum number as depicted in
Fig.\ref{coupling}. Analogously to the treatment of pion
condensation, the lowest-energy collective particle-hole mode is
interpreted as an effective vector meson field operating on the
ground state of the nucleus, i.e.,
\be
\frac{1}{\sqrt{A}}\sum_{i}[N^*_iN_i^{-1}]^{1-} \simeq \sum_{i}
[\rho(x_i)~ or~ \omega(x_i) ]|\Psi_0>_s,\label{col}
 \ee
 with the
antisymmetrical (symmetrical) sum over neutrons and protons giving
the $\rho$-like ($\omega$-like) nuclear excitation. Here the
``particle" is taken to be $N^\star$ while the ``hole" is
nucleon-hole. We will ignore the nucleon as particle since in the
channel we are concerned with, we expect the nucleon to be more
weakly coupled than the $N^\star$ to the (near on-shell) vector
meson. We call the collective mode (\ref{col}) ``sobar," i.e.,
$\rho$-sobar, $\omega$-sobar etc.

The dropping vector meson masses could then be calculated in terms
of mixing of the nuclear collective state, Eq.(\ref{col}), with
the elementary vector meson through the mixing matrix elements of
Fig.\ref{coupling}. Now building up the collective ``nuclear
mode," the latter can be identified as an analog to the state in
the degenerate schematic model of Brown for giant dipole
resonance~\cite{gerry}. The fields figuring in a BR scaling chiral
Lagrangian are then to be identified with interpolating fields for
the lowest branch modes that emerge from the mixing. An important
development which leads to the assumption Eq.(\ref{col}) was
furnished by Friman, Lutz and Wolf\cite{friman}. From empirical
values of the amplitudes such as $\pi + N\rightarrow \rho + N$
etc. they constructed the $\rho$-like or $\omega$-like states in
consistency with our assumption Eq.(\ref{col}). Thus the input
assumption made for the sobar model receives substantial empirical
support.

Since the development is at its initial stage and still quite
crude, we briefly summarize what we hope to accomplish in the end.

The property of a vector meson, say, $\omega$, in medium is
encoded in the propagator of the meson in interaction with the
system. For simplicity of discussion, let us consider a two-level
schematic model. In (\ref{col}), we take only one configuration
with $N^\star=N^\star (1520)$ in the $\omega$ channel. The
starting point is the $\omega$-meson propagator in nuclear matter
given by
\be
D_\omega(q_0,\vec q;\rho_N)=\frac{1}{q_0^2-\vec q^2 -m_\omega^{2}
-\Sigma_{\omega N^* N}(q_0,q;\rho_N)} \label{rhop1} \ee where we
have ignored the $\omega$ decay width, and the density-independent
real part of the self-energy has been absorbed into the free
(physical) mass $m_\omega$. Here $\rho_N$ is nucleon number
density. Note that within the low-order approximation made here
the entire density-dependence resides in the in-medium $\omega$
self-energy $\Sigma_{\omega N^* N}$ induced by $N^*(1520)N^{-1}$
excitations. In what follows we will for simplicity concentrate on
the limit of vanishing three-momentum where the longitudinal and
transverse polarization components become identical. Due to the
rather high excitation energy of $\Delta E=M_{N^*}-M_N=580$~MeV,
one can safely neglect nuclear Fermi motion to obtain
\be
\Sigma_{\omega N^* N}(q_0)\sim g_{\omega N^* N}^2
\frac{q_0^2}{m_\omega^2}\frac{\rho_N}{4} \left(\frac{2(\Delta
E)}{(q_0+i\Gamma_{tot}/2)^2-(\Delta E)^2}\right) \label{RWself}
\ee where $\Gamma_{tot}$ is the sum of the full width of $N^\star
(1520)$ in free space and density-dependent width due to medium.
If the widths of the $\omega$ and $N^*(1520)$ are sufficiently
small one can invoke the mean-field approximation and determine
the quasiparticle excitation energies from the zeros in the real
part of the inverse propagator. In particular, for $\vec q=0$, the
in-medium $\omega$ mass is obtained by solving the dispersion
relation
 \be q_0^2
=m_\omega^2+{\rm Re} \Sigma_{\omega N^* N}(q_0) \ . \label{disp}
 \ee
The pertinent spectral weights of the solutions are characterized
by $Z$-factors defined through
\be
Z=(1-\frac{\partial}{\partial q_0^2}{\rm Re} \Sigma_{\omega N^*
N})^{-1} \ . \label{Zfac} \ee

The formulas written above are presumably valid for low density
since they can be made consistent (by fiat) with low-energy
theorems. However there is no reason to expect that a low-order
calculation in strong coupling will be viable at high density. For
instance there is no way that the $\omega$ mass will go to zero at
any density even in the chiral limit. We are therefore led to make
certain assumptions motivated by our {\it objective} to model BR
scaling. It is clear that with a few-order perturbative
calculation in a strong-coupling regime, {\it there is no way to
arrive at BR scaling.} Lacking a workable scheme to compute
systematically, we will simply {\it impose} a condition on the
model and study the consequence on the model. The simplest
condition that we can impose is that $q_0=0$ be a solution of
(\ref{disp}) at some density $\rho_c$ at which the in-medium pion
decay constant $f_\pi^\star$ is to vanish (a la, e.g., in-medium
Weinberg sum rule) . This is readily achieved if
 \be {g_{\omega
N^* N}^\star}^2 \frac{q_0^2}{m_\omega^2}\rightarrow {\rm
constant}\label{limit} \ee
 independent of density as $\rho\rightarrow \rho_c$. Note that we
have appended a star on the $\omega N^* N$ coupling constant to
indicate that higher order corrections will inject a non-linear
density dependence into the vertex (as well as into the width
etc.) The limit can be achieved only if the density dependence in
$f^\star$ cancels the same in $q_0$ as one approaches the
critical density. Now the constant cannot be fixed a priori and
what one takes for it will determine at what $\rho_c$ the
effective $\omega$-sobar mass will vanish. The basic assumption
here is that since the vector mass drops while the pion mass does
not, the quasiparticle picture gets {\it restored} as $\rho$
approaches $\rho_c$ with the width shrinking due to the
decreasing phase space. This is consistent with the general
premise of BR scaling.

As stressed in \cite{sobar}, nobody has been able to ``derive"
such a sobar description starting from effective field theories
defined at zero density. It seems however promising that this is
doable in a systematic way in the framework laid down in
\cite{sobar}. How this can come about is sketched in the
references \cite{sobar}.
\subsection*{Acknowledgments}
\itt I am very grateful for discussions and collaborations with
G.E. Brown, B. Friman, Y. Kim, K. Kubodera, D.-P. Min, T.-S.
Park, and Chaejun Song. I would particularly thank B. Friman for
his insightful advice and comments for this article. Part of this
paper was written while I was visiting Korea Institute for
Advanced Study whose hospitality is acknowledged.

\end{document}